\documentclass[11pt]{article}
\usepackage{geometry}                
\usepackage{graphicx}
\usepackage{amssymb}
\usepackage{epstopdf}
\usepackage{amsmath, amsthm}
\usepackage{longtable}
\newtheorem{theorem}{Theorem}

\newtheorem{lemma}[theorem]{Lemma}

\newcommand{\bulk}{\mathrm{bulk}}
\newcommand{\soft}{\mathrm{soft}}
\newcommand{\hard}{\mathrm{hard}}

\begin{document}

\title{Asymptotic forms for hard and soft edge general $\beta$ conditional gap probabilities }
\author{Peter J. Forrester and Nicholas S. Witte}
\date{}
\maketitle

\noindent
\thanks{\small Department of Mathematics and Statistics, 
The University of Melbourne,
Victoria 3010, Australia email:  p.forrester@ms.unimelb.edu.au  \: nsw@ms.unimelb.edu.au
}

\begin{abstract}
\noindent  An infinite log-gas formalism, due to Dyson, and independently Fogler and Shklovskii, is applied to the computation
of conditioned gap probabilities at the hard and soft edges of random matrix $\beta$-ensembles. The conditioning is that there are
$n$ eigenvalues in the gap, with $n \ll |t|$, $t$ denoting the end point of the gap. 
It is found that the entropy term in the formalism must be replaced by a term involving the potential
drop to obtain results consist with known asymptotic expansions in the case $n=0$.
With this modification made for general $n$, the derived
expansions --- which are for the
logarithm of the gap probabilities --- are conjectured to be correct up to and including terms O$(\log|t|)$. They are shown to satisfy
various consistency conditions, including an asymptotic duality formula  relating $\beta$ to $4/\beta$.
\end{abstract}

\section{Introduction}

Fundamental to Dyson's \cite{Dy62e} pioneering studies of random matrices is the log-gas picture. 
The essential point is that the joint eigenvalue probability density function for various random matrix ensembles can be computed exactly, 
and is seen to be identical to the Boltzmann factor for certain classical particle systems interacting via a repulsive logarithmic pair potential. 
These so-called log-gas systems are in equilibrium at particular values of the dimensionless inverse temperature $\beta$, 
restricted to the value $\beta=1$ in the case of the matrix ensemble exhibiting an orthogonal symmetry, 
$\beta=2$ for a unitary symmetry and $\beta=4$ for a unitary symplectic symmetry. In the application of random matrix ensembles to 
quantum mechanics these cases correspond to a time reversal symmetry $T$ with $T^2 =1$, no time reversal symmetry, and a time reversal 
symmetry $T$ with $T^2= -1$, respectively.

The significance of the log-gas picture is that it permits the use of macroscopic arguments to compute certain probabilistic 
quantities in the bulk scaling limit (where the number of eigenvalues goes to infinity with their bulk density fixed to the value $\rho$). 
Specifically, the particle system is approximated as a continuum fluid confined to a one-dimensional domain but obeying the laws of two-dimensional electrostatics.
As an example, consider $E^{\bulk}_{\beta} (0; (0,t))$, the probability that there are no eigenvalues in the interval $(0,t)$ 
for the bulk scaling limit of a random matrix ensemble indexed by $\beta$. Dyson's first application of the log-gas analogy was to derive the asymptotic form
\begin{align}
\label{Eb} \log E^{\bulk}_{\beta} (0; (0,t)) \mathop{\sim}\limits_{t \to
\infty} -\beta \frac{(\pi \rho t)^2} {16} +\left( \frac{\beta} {2}- 1 \right)\frac{\pi \rho t} {2} +{\rm O}(\log \rho t).
\end{align}
Although it was remarked \cite[above eq.~(92)]{Dy62e} that the variational principle underlying (\ref{Eb}) may contain errors of order $\log \rho t$, 
other arguments were presented that suggested the log-gas prediction of the explicit term
\begin{align}
\label{Eb1} \frac{(1- \beta/2)^2}{2\beta} \log \rho t ,
\end{align}
may in fact be correct at this order.

A noteworthy feature is that (\ref{Eb}) is valid for all $\beta>0$, and thus in particular applies uniformly to all three random matrix symmetries. 
In fact, the general $\beta>0$ form of (\ref{Eb}) is relevant to contemporary random matrix theory \cite{Fo10}. Thus there are random tridiagonal 
matrices \cite{DE02} and random Hessenberg unitary matrices \cite{KN04} that $\beta$-generalize the eigenvalue probability density functions for the 
Gaussian and circular ensembles from classical random matrix theory. A remarkable development relating to these constructions has been the characterization of 
$E^{\bulk}_{\beta} (0; (0,t))$ in terms of a stochastic differential equation \cite{KS06}, \cite{VV07}. This was subsequently used \cite{VV08} 
to give a rigorous derivation of (\ref{Eb}), and to furthermore establish that the explicit form of the  O$(\log \rho t)$ term is not (\ref{Eb1}) but rather
\begin{align}
\label{Eb2} \frac{1}{4} \Big ( \frac{\beta}{2} + \frac{2}{\beta} - 3 \Big ) \log \rho t.
\end{align}

Generalizing $E^{\bulk}_{\beta} (0; (0,t))$ is the probability $E^{\bulk}_{\beta} (n; (0,t))$ of there being exactly $n$ eigenvalues in the interval 
$(0,t)$ for the bulk scaling limit of a random matrix ensemble indexed by $\beta$. Many years after his pioneering work \cite{Dy62e}, Dyson \cite{Dy95} and 
independently Fogler and Shklovskii \cite{FS95} used the log-gas picture --- now applied directly to the infinite bulk state (in  \cite{Dy62e} use was made
of a scaling limit of a large deviation formula in the finite system; see \S \ref{sVs}) --- to derive the generalization of (\ref{Eb})
\begin{multline}\label{Eb3}
\log E_{\beta}^{\bulk} (n; (0,t)) \mathop{\sim}\limits_{t,n\to \infty \atop t\gg n} 
  -\beta \frac{(\pi \rho t)^2} {16} + \left(\beta n +\frac{\beta} {2}- 1\right) \frac{\pi \rho t}{2}\\
  +\frac{n}{2} \left( 1- \frac{\beta}{2} -\frac{\beta n}{2} \right) \left( \log \frac{4\pi \rho t}{n} +1\right).
\end{multline}
In fact, in the cases $\beta=1, 2$ and $4$ this same formula, except requiring $n$ to be fixed, had been obtained earlier \cite{BTW92} 
using an analysis based on explicit Painlev\'{e}/Fredholm determinant expression for $E^{\bulk}_{\beta} (0; (0,t))$.

The primary aim of the present paper is to extend Dyson's \cite{Dy95} log-gas study of the bulk conditioned gap probability
$E^{\bulk}_{\beta} (n; (0,t))$ to the analogous quantities $E^{\soft}_{\beta} (n; (t,\infty))$ and $E^{\hard}_{\beta} (n; (0,t))$ corresponding 
to there being exactly $n$ eigenvalues in the intervals $(t,\infty)$ and $(0,t)$ about the soft and hard spectrum edges respectively, for the regime $1 \ll n \ll |t|$, $|t| \to \infty$.
The soft edge refers to the neighbourhood of the largest or smallest eigenvalue, in the situation that the corresponding eigenvalue density 
exhibits a square root profile, which from our formalism extending \cite{Dy95}  is taken to be (semi) infinite in extent.  The hard edge is the neighbourhood of the smallest eigenvalue for matrices with non-negative 
eigenvalues, when the spectral density exhibits an inverse square root singularity profile, which again is to be taken of (semi) infinite  extent.

As examples of explicit realizations of these edge states in terms of limits of finite $N$ matrix ensembles, let ME${}_{\beta,N}(g)$ denote the eigenvalue probability density 
 proportional to 
 \begin{align}
\nonumber \prod_{l=1}^N g(\lambda_l) \prod_{1\leq j < k\leq N} |\lambda_k - \lambda_j|^{\beta} .
\end{align}
The choice $g(\lambda) = e^{- \beta \lambda^2/2}$ defines the Gaussian $\beta$-ensemble, and the choice $g(\lambda) = \lambda^{\beta a/2} e^{-\beta \lambda /2} \,(\lambda>0)$ defines the  Laguerre $\beta$-ensemble. We know from \cite{Fo93a} that the largest eigenvalues are, respectively, to leading order in $N$
at $\lambda = \sqrt{2N}$ and $\lambda = 4N$, and that  \cite{ES06,RRV06} upon the scalings  \cite{Fo93a}
$$
 \lambda_l = \sqrt{2 N} + \frac{x_l} {\sqrt{2}\; N^{1/6}} , \qquad  \lambda_l = 4N +2(2N)^{1/2} x_l ,
 $$
the correlations have the same well defined $N \to \infty$.
The corresponding scaled density $\rho_{(1)}^{\soft, \beta} (x)$ admits an explicit $\beta$-dimensional integral formula in the case of $\beta$ even \cite{DF06},
and this exhibits the square root singularity characterizing the soft edge state by way of the asymptotic formula
\begin{align}\label{softd}
 \rho_{(1)}^{\soft, \beta} (x) \mathop{\sim}\limits_{x\to -\infty} \frac{\sqrt{|x|}}{\pi}.
\end{align}
The standard realization of the hard edge is the Laguerre $\beta$-ensemble, upon the scaling 
\cite{Ed88,Fo93a}
\begin{align}
\nonumber \lambda_l = \frac{x_l} {4N}.
\end{align}
For $a=c-2/\beta$, $c$ a positive integer, and $\beta$ even the corresponding scaled density $\rho_{(1)}^{\hard, \beta} (x)$ 
admits an explicit $\beta$-dimensional integral formula \cite{Fo94b}. Asymptotic analysis of this integral shows
\begin{align}\label{hardd}
 \rho_{(1)}^{\hard, \beta} (x) \mathop{\sim}\limits_{x\to \infty} \frac{1}{2\pi \sqrt{x}},
\end{align}
thus exhibiting the characterizing density profile at the hard edge.

The relevance of the spectrum edge to problems in theoretical and mathematical physics was not apparent in the era of Dyson's first works.
In fact, at the soft edge there is a somewhat hidden, but nonetheless fundamental, relationship to growth problems of the KPZ class 
(see e.g. the recent review \cite{FS11}). And in the case of the hard edge the statistical properties of the smallest eigenvalue can be 
related to data from lattice QCD simulations (see e.g.~\cite{Ve11}).

A priori the log-gas formalism can be applied to the study of the asymptotic form of gap probabilities in the cases the gap contains no eigenvalues,
or in the cases there are $n$ eigenvalues with $1 \ll n \ll |t|$. In the former case the asymptotic form of the logarithm of the gap probability for general
$\beta$ at both the hard and soft edges is known from other considerations \cite{Fo94b,CM94,BEMN10,RRZ11}, up to and including terms O$(\log |t|)$.
This information is important for two reasons. First, it tells us the infinite log-gas formalism at 
both the hard and soft edge is in error in the second order (free energy/ entropy) contribution.
But fortunately, guided by the bulk case, an ansatz can be made to correct this error. Second, it allows the asymptotic expansions obtained
for $1 \ll n \ll |t|$ to be very naturally extended to hold (as a conjecture) uniformly in $n$, up to and including terms
O$(\log |t|)$.

These latter expansions, which are the main result of the paper, read
\begin{multline}\label{22a}
\log E_\beta^{\rm hard}(n;(0,t)) \mathop{\sim}\limits_{t \to \infty \atop n \ll t}
- \beta \bigg\{ \frac{t}{8} - \sqrt{t} \left(n + \frac{a}{2} \right) \\
+ \Big[ \frac{n^2}{2} + \frac{na}{2} + \frac{a(a-1)}{4} + \frac{a}{2\beta} \Big] \log t^{1/2} \bigg\},
\end{multline}
and
\begin{multline}\label{31h}
\log E_\beta^{\rm soft}(n;(t,\infty)) \mathop{\sim}\limits_{t \to - \infty \atop n \ll |t|} 
 - \frac{\beta|t|^3}{24} + \frac{\sqrt{2}}{3} |t|^{3/2} \left(\beta n + \frac{\beta}{2} -1\right)
 \\
 + \left[ \frac{\beta}{2} n^2 + \left( \frac{\beta}{2} - 1 \right) n + \frac{1}{6}  \left( 1 - \frac{2}{\beta} \left(1 - \frac{\beta}{2} \right)^2 \right) \right]\log |t|^{-3/4}.
\end{multline}
As a check, using results from \cite{Fo09}, asymptotic duality equations --- relating gap probabilities with given $\beta,n$ to gap probabilities at
$4/\beta$ and with modified $n$ and length scale--- are identified for both $E_\beta^{\rm hard}(n;(0,t))$ and
$ E_\beta^{\rm soft}(n;(t,\infty))$, and (\ref{22a}) and (\ref{31h}) are shown to exhibit the required inter-relations.

\section{Strategy of the log-gas argument}
\setcounter{equation}{0}
\subsection{Formalism for a general background potential}

The finite log-gas system consists of $N$ mobile charges, of strength $+1$ say, and a smeared out neutralizing background charge density $-\rho_b (x)$. 
We are interested in the $N\to \infty$ limit, in the circumstance that $\rho_b(x)$ is supported on some semi-infinite or infinite domain $I$. 
A physical argument tells us that an equilibrium state will exhibit local charge neutrality, and so $\rho_{(1)}(x)=\rho_b(x)$, 
where $\rho_{(1)}(x)$ denotes the partial density (otherwise the charge imbalance would create a non-zero electric field, 
which in turn would mean the system is not in equilibrium).

Let $E_{\beta} (n; J)[\rho_b]$ denote the probability that in this system there are precisely $n$ eigenvalues in the interval $J$. 
The basic hypothesis (see e.g. \cite[\S 14.6.2]{Fo10}) underlying the log-gas argument is that for $|J|\to \infty$
\begin{align}
\label{D6} \log E_{\beta} (n; J)[\rho_b] = -\beta \delta F,
\end{align}
where $\delta F$ is the change in energy resulting from conditioning the equilibrium particle density so that the interval $J$ contains $n$ particles. 
Furthermore, the change in energy consists of two distinct parts, namely the electrostatic energy $V_1$ and the free energy $V_2$, 
the latter being entirely due to entropy.

The electromagnetic energy is responsible for the leading order (in $|J|$) contribution to (\ref{D6}). With $\rho_{(1)}(x)$ 
now denoting the new equilibrium particle density, and
\begin{align}
\label{Vp} \phi (x) = -\int_{I} (\rho_{(1)} (y) - \rho_{b}(y)) \log |x-y| dy ,
\end{align}
denoting the electrostatic potential at a point $x$ due to the perturbed charge density, we have
\begin{align}
\label{V1} V_1 = \frac{1}{2} \int_{I} (\rho_{(1)} (x) - \rho_{b}(x)) \phi(x) dx.
\end{align}
The second-order contribution to (\ref{D6}) results from both the electrostatic energy and the free energy. Let $f_{\beta}= f_{\beta} [\rho_{(1)} (x)]$ 
denote the density function for the free energy per particle at point $x$. The change in this density function for the free energy per particle due to the perturbation is
\begin{align}
\nonumber f_{\beta} [\rho_{(1)} (x)] -f_{\beta} [\rho_{b} (x)] = \left(\frac{1} {\beta} - \frac{1}{2} \right) \log \left( \frac{\rho_{(1)}(x)} {\rho_b (x)}\right),
\end{align}
where the equality follows from the exact expression for $f_{\beta}[\rho]$ (see e.g. \cite[\S 4.8.1]{Fo10}). Thus the total change in free energy  is
\begin{align}
\label{V2} V_2 = \left( \frac{1}{\beta} - \frac{1}{2}\right) \int_{I} \rho_{(1)} (x) \log \left( \frac{\rho_{(1)}(x)} {\rho_b (x)}\right) dx.
\end{align}

With the background charge density $\rho_{b} (x)$ given, two-dimensional electrostatics is to be used to determine the complex electric field
\begin{align}
\label{Ez} E(z) = \int_I \frac{\rho_{(1)}(x) -\rho_b (x)} {z-x} dx,
\end{align}
which in turn is used to compute the particle density $\rho_{(1)} (x)$ and electrostatic potential $\phi(x)$. The validity of this approach is restricted to large 
gap size $|J|$. Thus it is only in this circumstance that we are able to use physical principles to fix $\phi(x)$ in certain domains and, 
furthermore, to specify some properties of $\rho_{(1)} (x)$ to the extent that $E(z)$ can be uniquely determined.

To see this suppose for definiteness that $1\ll n \ll |J|$. We then expect the $n$ particles to form a macroscopic blob within some interval $J' \subset J$, and thus
\begin{align}
\label{d8} \rho_{(1)} (x)=0, \quad \mathrm{for} \; x\in J\backslash J' ,
\end{align}
while
\begin{align}
\label{d8o} \int_{J'} \rho_{(1)} (x) dx = n.
\end{align}
Furthermore, we expect $\rho_{(1)} (x)$ to vanish as $x$ approaches the boundary of $J'$ from within $J'$ (a soft edge singularity), 
while we expect $\rho_{(1)} (x)$ to diverge as $x$ approaches the boundary of $J$ from within $J$ (a hard edge singularity). 
It could also happen that the left hand end points of $J$ and $J'$ coincide, in which case a hard edge singularity of $\rho_{(1)} (x)$ is expected.

From the viewpoint of macroscopic electrostatics the regions $J'$ and $I\backslash J$, being occupied by what is effectively a continuum of mobile charges, 
behave as perfect conductors and thus have $\phi (x)$ equal to a constant throughout. As a normalization, 
in the (semi) infinite region $I\backslash J$ this constant can be chosen to equal zero. Then for $x\in J'$ we will have $\phi (x)=-v$,
 with $v>0$. Furthermore, $\phi (x)$ must vary continuously from this value to zero, and, with $t$ the smallest end point of $I\backslash J$ and $J'_b$ the largest end point of $J'$,
\begin{align}
\label{thp} \int_{J'_b}^t \frac{d\phi} {dx} dx = v.
\end{align}

The fact that the particle density vanishes in the complement of the regions of constant potential allows (\ref{V1}) to be simplified to read
\begin{align}
\nonumber V_1 & = \frac{v}{2} \int_{J'} (\rho_{(1)}(x)- \rho_b(x)) dx -\frac{1} {2} \int_{J\backslash J'} \rho_b (x) \phi (x) dx\\
\label{th1} &=-\frac{vn} {2} - \frac{1} {2} \int_{J} \rho_b (x) \phi (x) dx,
\end{align}
where to obtain the second equality use has been made of (\ref{d8o}). And with the left hand end point of $J$ denoted $J_a$, 
integration by parts together with the fact that $\phi (J_a)=0$ can be used to conclude from (\ref{th1}) that
\begin{align}
\label{V1s} V_1 = -\frac{vn}{2} +\frac{1}{2} \int_{J\backslash J'} \left( \int_{J_a}^x \rho_b (y) dy\right) \frac{d\phi}{dx}dx.
\end{align}
Since $\rho_b(y)$ is given, we see from (\ref{V1s}) and (\ref{thp}) that $V_1$ depends only on $\frac{d\phi} {dx}$, 
and not on both $\rho_{(1)} (x)$ and $\phi (x)$ as appears in (\ref{V1}). But knowledge of $\rho_{(1)} (x)$ is required for the computation of $V_2$ as specified by (\ref{V2}).

\subsection{The background density $\rho_b (x)= cx^{\alpha}$ with $n=0$}

We now specialize to a semi-infinite domain $I=(0, \infty)$ and background density $\rho_b (x)= cx^{\alpha}, \alpha >-1$. 
This includes as special cases the soft edge ($\alpha = 1/2$; recall (\ref{softd})) and hard edge ($\alpha = -1/2$; recall (\ref{hardd})).
It also includes the density profiles of the so-called multicritical matrix ensembles, for which $\alpha = 2k +1/2$, $k \in \mathbb Z_{\ge 0}$; 
see e.g.~\cite{CIK09}.
Let $J=(0,t)$ be conditioned to have precisely $n$ eigenvalues. Although in the above discussion it has been assumed $n\gg 1$, 
repeating the reasoning for the case $n=0$ shows that the appropriate limit of (\ref{V1s}) remains valid, and thus
\begin{align}
\nonumber V_1&= \frac{1}{2} \int_{J} \left( \int_0^x \rho_b(y) dy\right) \frac{d\phi}{dx} dx\\
\label{V1t} &= \frac{c}{2(\alpha+1)} \int_0^t x^{\alpha+1} \frac{d\phi}{dx}dx.
\end{align}

In this special case the particle density $\rho_{(1)}(x)$ depends only on $x$ and $t$. Moreover, as observed in \cite{Fo93a}, 
the $t$ dependence can be factorized by the scaling ansatz
\begin{align}
\label{V1u} \rho_{(1)}(tx) = t^{\alpha} \left( \rho_{(1)}(x)\Big|_{t=1} \right),
\end{align}
which is, of course, a property of $\rho_b (x)$. Assuming (\ref{V1u}), the complex electric field (\ref{Ez}) exhibits an analogous scaling
\begin{align}
\nonumber E(tz) = t^{\alpha} \int_1^{\infty} \frac{\rho_{(1)}(x)\Big|_{t=1} -\rho_b (x)} {z-x} dx.
\end{align}
But
\begin{align}
\label{11.0} \frac{d\phi} {dx} = -\lim_{y\to 0^{+}} \mathrm{Re} \:E(z) ,
\end{align}
and so
\begin{align}
\left. \frac{d\phi} {dx} \right|_{x\mapsto xt} = t^{\alpha} \left.\frac{d\phi} {dx}\right|_{t=1}.
\end{align}
Making use of this latter formula in (\ref{V1t}) we obtain
\begin{align}
\label{11.1} V_1= \tilde{c}\: t^{2\alpha+2}, \quad \tilde{c} = \frac{c}{2(\alpha+1)}\int_0^1 x^{\alpha+1} \left.\frac{d\phi} {dx}\right|_{t=1} dx,
\end{align}
and thus, upon substituting in (\ref{D6}) for $\delta F$, the prediction \cite{Fo93a}  that for $t\to \infty$
\begin{align}\label{2.15ax}
 \log E_{\beta} (n; (0,t)) [c x^{\alpha}] \sim -\beta \: \tilde{c}\: t^{2(\alpha +1)}.
\end{align}

To go beyond this scaling prediction, and compute for example the explicit form of $\tilde{c}$ in 
(\ref{11.1}), we seek to deduce the explicit functional form of $E(z)$ as a consequence of its necessary properties. 
In relation to the latter, we begin by noting that the charge neutrality condition
\begin{align}
\nonumber \int_I (\rho_{(1)}(x) -\rho_b(y)) dy =0 ,
\end{align}
implies
\begin{equation}
\label{Ezx} E(z) \mathop{\sim}\limits_{|z|\to\infty} {\rm o}(z^{-1}).
\end{equation}
Another condition is that as $t\to \infty$ we require
\begin{align}
\label{Ez2} E(z)\to -c\int_0^{\infty} \frac{x^{\alpha}} {z-x}dx,
\end{align}
which is the electric field due to the background (this is well defined for $-1< \alpha< 0$; for other $\alpha>0$ it is specified by analytic continuation).

By considering the contour integral
\begin{align}
\nonumber \int_{\Sigma} \frac{w^{\alpha}} {z-w} dw ,
\end{align}
where $\Sigma$ is the contour traversing the ray $\theta = \arg z$ from infinity down to the origin, indented at the point $z$, 
then the positive real axis, we can evaluate (\ref{Ez2}) to obtain
\begin{align}
E(z) &\to -c z^{\alpha}\; \mathrm{PV} \int_0^{\infty} \frac{x^{\alpha}} {1-x} dx +\pi i cz^{\alpha}\\
&= cz^{\alpha} \left( -\mathrm{PV} \int_{0}^{\infty} \frac{x^{\alpha}} {1-x} dx +\pi i \right) ,
\end{align}
where $\mathrm{PV}$ denotes the principal value. By breaking the integral at $x=1$ and changing variables $x\mapsto 1/x$ in the range $x\in (1,\infty)$ 
the principal value integral can be evaluated to give
\begin{align}
\label{Ez3} E(z) \to cz^{\alpha} (-\pi \cot \pi \alpha +\pi i).
\end{align}
We see immediately that the values $\alpha = (k-1/2)$, $k=0,1, \dots$ are distinguished, as then $E(z)$ is pure imaginary, and
\begin{align}
\label{Ez4} E(z) \to \pi i c z^{\alpha} \qquad \mathrm{as}\; t\to \infty.
\end{align}
This turns out to be crucial in constructing $E(z)$ so that, for example, (\ref{11.0}) vanishes for $x>t$, and so restriction to these values of $\alpha$ will be assumed henceforth.

To continue, we use the fact that the only singularity of $E(z)$, apart from $z=0$ as already present in $\rho_b (z)$, is $z=t$.
Furthermore, this is required to be an inverse square root, corresponding to an (induced) hard wall singularity. We see immediately that for $\alpha=-1/2$
\begin{align}
\label{13.1} E(z) = i \pi c \frac{1}{\sqrt{z}}\left( 1-\sqrt{\frac{z} {z-t}} \right) ,
\end{align}
has this property, as well as the necessary properties (\ref{Ez4}) and (\ref{Ezx}). Furthermore, the real part of $E(z)$ on the real axis vanishes for $x>t$, 
telling us via (\ref{11.0}) that the potential is a constant in this region. And the general formula
\begin{align}
\label{13.2} \rho_{(1)}(x) - \rho_{b} (x) = -\frac{1}{\pi} \hspace{-6pt} \lim_{\quad y\to 0^{+}} \mathrm{Im}\; E(z),
\end{align}
which follows from (\ref{Ez}) when applied to (\ref{13.1}) gives $\rho_{(1)} (x)=0$ for $0<x<t$. Hence, we conclude that (\ref{13.1}) 
satisfies all necessary conditions required from physical principles.

We now proceed to use (\ref{13.1}) to compute $V_1$. Thus substituting in (\ref{11.0}) gives that for $0<x<t$
\begin{align}
\nonumber \frac{d\phi} {dx} = \frac{\pi c} {\sqrt{t-x}}.
\end{align}
Substituting in (\ref{V1t}), scaling the $t$ dependence and computing the integral (a special case of the Euler beta integral) gives
\begin{align}
\label{13.2a} V_1 = \frac{(\pi c)^2} {2} t,
\end{align}
 thus specifying $\tilde{c}$ in the case $\alpha = -1/2$.

For $\alpha = k-1/2$ with $k\in \mathbb{Z}^{+}$ we generalize (\ref{13.1}) to read
\begin{align}
\label{14.1} E(z) = \pi i c z^{\alpha} \left( 1- \sqrt{\frac{z} {z-t}} p_k(tz^{-1}) \right) ,
\end{align}
where
\begin{align}
\nonumber p_k(x) &= \sum_{j=0}^k \frac{x^j} {j!} \left. \left( \frac{d^j}{dx^j}(1-x)^{1/2} \right) \right|_{x=0}\\
\nonumber &= \sum_{j=0}^k \frac{(-\tfrac{1}{2})_j \; x^j} {j!}, \qquad \qquad (u)_j:= u(u+1)\cdots (u+j-1).
\end{align}
This modification is needed for the necessary condition (\ref{Ezx}) to be satisfied, while leaving unchanged the validity of the other necessary conditions. 
It tells us that for $0<x<t$
\begin{align}
\frac{d\phi} {dx} = \pi c \frac{z^k} {\sqrt{z-t}} p_k(tz^{-1}),
\end{align}
and this in turn substituted in (\ref{V1t}) implies
\begin{align}
\label{14.2} V_1= \frac{\pi c^2} {2k+1} t^{2k+1} \sum_{j=0}^k \frac{\Gamma (2k-j +3/2) \Gamma (1/2)}{\Gamma (2k-j+2)} \frac{(-1/2)_j} {j!}.
\end{align}
The simplest case is $k=1$, corresponding to $\rho_b (x)= c \sqrt{x}$, for which (\ref{14.2}) simplifies to read
\begin{align}
\label{14.2a} V_1 = \frac{\pi^2 c^2} {24} t^3.
\end{align}

In the cases $\alpha=\pm 1/2$, the background density $\rho_b (x) =c x^{\alpha}$ has distinguished values of the proportionality constant
\begin{align}
\label{14.3} c= 
\begin{cases}
\frac{1}{2\pi}, & \alpha =-\frac{1}{2}\\
\frac{1}{\pi}, & \alpha =\frac{1}{2}
\end{cases} .
\end{align}
These are read off from (\ref{hardd}) and (\ref{softd}) respectively.
Substituting the values of $c$ from (\ref{14.3}) in (\ref{13.2a}) and (\ref{14.2a}) respectively, 
then substituting the corresponding values of $V_1$ for $\delta F$ in (\ref{D6}) with $n=0$ predicts that
\begin{align}
\label{16.1} &\log E_{\beta}^{\soft} (0;(t, \infty)) \mathop{\sim}\limits_{t\to -\infty} -\frac{\beta}{24} |t|^3,\\
\label{16.2} &\log E_{\beta}^{\hard} (0;(0,t)) \mathop{\sim}\limits_{t\to \infty} -\frac{\beta}{8} t.
\end{align}
The first of these predictions agrees with results of recent rigorous asymptotic analysis \cite{RRV06} based upon the characterization of 
$E_{\beta}^{\soft} (0;(t, \infty))$ in terms of stochastic differential equations \cite{ES06}; 
the second agrees with the result obtained from exact asymptotic analysis of an $m$-dimensional integral formula available for 
$E_{\beta}^{\hard} (0;(0,t))$ in the case that $\beta a/2 =m \in \mathbb{Z}_{\geq 0}, \beta>0$ general \cite{Fo94b}. 
We remark that (\ref{16.2}) has in fact been derived earlier in a log-gas analysis \cite{CM94}. The implementation though was different to that used here,
as it was applied to the hard edge of the finite $N$ Laguerre $\beta$-ensemble (see  \S \ref{sVs} below for further comments on finite $N$ strategies).

In the next two sections the log-gas argument leading to (\ref{16.1}) and (\ref{16.2}) will be refined to obtain the analogue of 
Dyson's asymptotic formulas for the bulk spacing (\ref{Eb}) and (\ref{Eb3}), for the soft and hard edges. We will begin with the latter, 
as it is then possible to make use of results already obtained in \cite{Dy95}. 
But before we proceed we remark that in the case of exponents $\alpha = 2k + 1/2$ (corresponding to the multicritical matrix ensembles)
and with $\beta = 2$, the asymptotic form
(\ref{2.15ax}) has recently been rigorously derived in \cite{CIK09}, together with an explicit evaluation of the proportionality. 
However the value of the latter
as implied by (\ref{14.2}) cannot readily be checked as the workings of  \cite{CIK09} do not make explicit the value of $c$ in the effective background
density $\rho_b(x) = c x^\alpha$.

\section{The hard edge}
\setcounter{equation}{0}

\subsection{Direct calculation}

In the log-gas formalism the hard edge is specified by the background density
\begin{align}
\nonumber \rho_b(x) = \frac{1}{2\pi \sqrt{x}},
\end{align}
and, furthermore, by an electrostatic coupling between a fixed charge of strength $a'$ at the origin, and the charge density,
\begin{align}
\label{17.2} V_1' := -a' \int_0^{\infty} \left( \rho_{(1)} (y) -\rho_b (y) \right) \log y \; dy,
\end{align} 
where
\begin{align}
\label{17.2a} a'= \frac{a-1}{2} +\frac{1}{\beta} ,
\end{align}
(see \cite{Fo10}). This additional term accounts for the parameter $a$ in the Laguerre weight $x^{a\beta /2} e^{-\beta x/2}$, which in the hard
edge scaling is independent of $N$.

Comparing (\ref{17.2}) with (\ref{Vp}) shows
\begin{align}
\label{17.3} V_1' = a' \phi (0).
\end{align}
Conditioning so that exactly $n$ eigenvalues lie in the interval $(0,t)$, and with $1\ll n \ll t$ we expect the $n$ particles to occupy $J'=(0,b), b\ll t$. 
This region will behave like a perfect conductor, and consequently $\phi(x)= -v, v>0$ and thus constant for $x\in (0,b)$. 
It follows from this that (\ref{17.3}) can be written
\begin{align}
\label{18} V_1'= -a'v.
\end{align}
Furthermore, in this setting (\ref{V1s}) reads
\begin{align}
\label{18a} V_1= -\frac{vn} {2} +\frac{1}{2\pi} \int_b^t \sqrt{x}\; \frac{d\phi}{dx} \; dx ,
\end{align}
and we have too that
\begin{align}
\label{18b} v=\int_b^t \frac{d\phi}{dx}\; dx.
\end{align}
Our immediate task is therefore to compute $d\phi/dx$ for $x\in (b,t)$.

For this purpose we must generalize (\ref{13.1}) to the present setting of a new soft edge singularity at $z=b$, 
while still upholding the other necessary conditions required of $E(z)$, as discussed above (\ref{13.1}). We see that the simple variation of (\ref{13.1})
\begin{align}
\label{18.1} E(z) = \frac{i}{2\sqrt{z}} \left(1- \sqrt{\frac{z-b}{z-t}} \right) ,
\end{align}
has these properties. This furthermore satisfies the requirement that its real part vanishes on the real axis for $x\in (0,b) \cup (t,\infty)$, 
as required by these regions being perfect conductors, while (\ref{13.2}) shows that $\rho_{(1)} (x)=0$ for $x\in (b,t)$. 
Thus all physical requirements are satisfied, so we can now use (\ref{18.1}) for predictive purposes.

In particular, for the task at hand, substituting (\ref{18.1}) in (\ref{11.0}) allows us to conclude that for $b<x<t$,
\begin{align}
\frac{d\phi} {dx} = \frac{1}{2}\sqrt{\frac{x-b}{x(t-x)}}.
\end{align}
Substituting in (\ref{18b}) and changing variables gives
\begin{align}
\label{19.1} v=\int_{\sqrt{b}}^{\sqrt{t}} dx\,\sqrt{\frac{x^2-b} {t-x^2}},
\end{align}
while substituting in (\ref{18a}) and simplifying gives
\begin{align}
\label{19.2} V_1= -\frac{vn}{2} + \frac{1}{8} (t-b).
\end{align}
The integral (\ref{19.1}), while not an elementary function of $t,b$ as in (\ref{19.2}), can be expressed in terms of the elliptic integrals $E', K'$ 
corresponding to the elliptic modulus 
\begin{equation}\label{kbt}
k=\sqrt{b/t}. 
\end{equation}
Thus we read off from \cite[eq. (2.32)]{Dy95} that
\begin{align}
\label{20.1a} v=\sqrt{t} \left[ E'-k^2 K' \right].
\end{align}
Insight into why elliptic functions appear in so-called `two-cut' (here $(0,b) \cup (t,\infty)$) log-potential electrostatic problems can
be found in e.g.~\cite{BDE00}

The endpoint $b$ is determined by setting $J'=(0,b)$ in (\ref{d8o}) and thus requiring
\begin{align}
\label{np} \int_0^b \rho_{(1)}(x) dx = n.
\end{align}
To determine $\rho_{(1)}(x)$ we substitute (\ref{18.1}) in (\ref{13.2}) and read off that for $0<x<b$,
\begin{align}
\label{20.2} \rho_{(1)} (x) = \frac{1}{2\pi} \sqrt{\frac{b-x} {x(t-x)}}.
\end{align}
Substituting in (\ref{np}) and changing variables $x\mapsto x^2$ allows, upon making use of \cite[eq. (2.31), (2.33)]{Dy95}, 
the integral to be written in terms of elliptic integrals $E,K$ of elliptic modulus $(\ref{kbt})$, giving
\begin{align}
\label{20.2a} n= \frac{\sqrt{t}}{\pi} \left[ E-(1-k^2) K \right].
\end{align}
From the elliptic function forms (\ref{20.1a}) and (\ref{20.2a}), and with knowledge of the expansions in \cite[eq. (2.36), (2.37)]{Dy95}, 
we deduce that for $t\to \infty$, and with $1\ll n \ll t^{1/2}$ and $k\ll 1$,
\begin{align}
\label{21.1} b^2 &= 4t^{1/2} n - 2n^2 + {\rm O}(n^3/t)\\
\label{21.2} v &= t^{1/2} - n\log t^{1/2} + {\rm O}(n).
\end{align}
Here, in obtaining (\ref{21.2}), use has been made of (\ref{21.1}). Substituting (\ref{21.2}) in (\ref{18}) and (\ref{19.2}) then shows that in this same limit
\begin{align}
\label{21.2a} V_1+V_1'=\frac{t}{8} - t^{1/2} (n+a') + \left( n^2/2 +a' n\right) \log t^{1/2} + {\rm O}(n^2).
\end{align}

It remains to compute (\ref{V2}). In addition to the explicit form of the density in $(0,b)$, the explicit form in its other non-zero domain $(t, \infty)$  is also required. 
But this is immediate from (\ref{18.1}) and (\ref{13.2}), which give
\begin{align}
\nonumber \rho_{(1)}(x) = \frac{1}{2\pi} \sqrt{\frac{x-b}{x(x-t)}} ,\qquad x\in (t,\infty),
\end{align}
and consequently we have
\begin{multline}
 V_2 = \frac{1}{\pi} \left( \frac{1}{\beta} - \frac{1}{2} \right) 
     \left( \int_0^{\sqrt{b}} dx\; \sqrt{\frac{b-x^2}{t-x^2}} \log \sqrt{\frac{b-x^2} {t-x^2}} \right.
\label{21.3} 
      +\left. \int_{\sqrt{t}}^{\infty} dx\;\sqrt{\frac{x^2-b}{x^2-t}} \log \sqrt{\frac{x^2-b}{x^2-t}} \right) ,
\end{multline}
(this form follows upon the change of variables $x\mapsto x^2$). An equivalent expression to this arises in Dyson's log-gas computation of the bulk spacing \cite{Dy95}, 
where it was evaluated in terms of the potential drop to give for the present expression
\begin{align}
\label{V2v} V_2 = \left( \frac{1}{\beta} - \frac{1} {2} \right)\frac{v}{2}.
\end{align} 
Its asymptotic form follows immediately upon substituting (\ref{21.2}).

The simplest case is $n=0$. With this assumed,
adding the asymptotic form of (\ref{V2v}) to (\ref{21.2a}) and recalling (\ref{17.2}) shows
$$
V_1 + V_1' +V_2 = \frac{t}{8} -\sqrt{t} \left[ \frac{a}{2} - \frac{1}{2}  \left( \frac{1}{\beta} - \frac{1}{2} \right) \right].
$$ 
Substituting this for $\delta F$ in (\ref{D6}) then gives the prediction
\begin{equation}\label{aj}
 \log E_{\beta}^{\hard} (0;(0,t))  \mathop{\sim}\limits_{t\to \infty} -\beta \left( \frac{t} {8} -k \sqrt{t} \right) ,
 \end{equation}
with $k =  \frac{a}{2} - \frac{1}{2}  \left( \frac{1}{\beta} - \frac{1} {2}  \right)$. But this contradicts
the elementary exact result  of \cite{Ed88,Fo94b} stating $\log E_{\beta}^{\hard} (0;(0,t))  |_{a = 0} = 
-\beta t/8$ .

It turns out that consistency with known exact results for $n=0$ can be obtained if we replace (\ref{V2v})
by
\begin{equation}\label{V2v1}
 V_2 = \left( \frac{1}{\beta} - \frac{1} {2} \right)v.
 \end{equation}
 This is precisely the evaluation of $V_2$ obtained from its definition (\ref{V2}) found by Dyson \cite{Dy95} in the bulk.
 Thus we then get (\ref{aj}) with $k = \frac{a}{2}$, in agreement with known exact results 
 from \cite{Fo94b, CM94,RRZ11}.
Hypothesizing (\ref{V2v1}) for general $n$ then gives
\begin{align}
\nonumber V_1 + V_1' +V_2 = \frac{t}{8} -\sqrt{t} \left(n+ \frac{a}{2} \right) + \left( \frac{n^2}{2} +\frac{an}{2} \right) \log t^{1/2} + {\rm O}(n^2).
\end{align}
Substituting this for $\delta F$ in (\ref{D6}), we then obtain the prediction that
\begin{align}
\label{22} \log E_{\beta}^{\hard} (n;(0,t))  
     \mathop{\mathop{\sim}\limits_{t\to \infty}}\limits_{1\ll n\ll t} -\beta \left\{ \frac{t} {8} -\sqrt{t} \left( n + \frac{a}{2} \right) 
          + \left( \frac{n^2} {2} + \frac{an} {2} \right) \log t^{1/2} \right\} .
\end{align}
One check on (\ref{22}) is that it is consistent with the inter-relation
\begin{align}
\label{22.1} E^{\hard}_{2} (n;(0,t)) 
\mathop{\mathop{\sim}\limits_{t\to \infty}}\limits_{n\ll t} E_1^{\hard} (n;(0,t))\Big|_{a \mapsto a-1} \; E_1^{\hard} (n+1;(0,t))\Big|_{a\mapsto a-1},
\end{align}
which in turn follows from \cite[eq. (3.5)]{Fo06c}.

A noteworthy feature of (\ref{22}) is that for $\beta=2$ it exhibits precisely the same functional form as the $t\to \infty$ expansion of 
$\log E_2^{\rm hard}(n;(0,t))$ with $n$ fixed \cite{TW94b}, \cite[eq.~(9.89)]{Fo10} up to terms $O (\log t)$ independent of $n$. 
In the case $n=0$ the exact form of the term $O(\log t)$ in the $t\to \infty$ expansion of $\log E_{\beta}^{\rm hard} (n;(0,t))$ 
is known for general $\beta$ \cite{Fo94b,CM94,RRZ11}. Adding this to (\ref{22}) suggests that uniformly in $n$, for $n\ll t$, 
the expansion (\ref{22a}) recorded in the Introduction is valid.

The generalized form (\ref{22a}) can be checked against the large $t$ duality formula 
\begin{align}
\label{23x} E_{\beta}^{\hard} (n; (0,t/ \tilde{s}_{\beta})) \mathop{\mathop{\sim}\limits_{t\to \infty}}\limits_{n\ll t} 
 E_{4/\beta}^{\hard} \left( \tfrac{1}{2}\beta(n+1)-1; \left( 0,t/\tilde{s}_{4/\beta} \right) \right)\Bigg|_{a\mapsto \beta a/2 -\beta +2},
\end{align}
 where $\tilde{s}_{\beta}$ is an arbitrary length scale that satisfies $\tilde{s}_{4/ \beta} (\beta/2)^2 = \tilde{s}_{\beta}$. 
This asymptotic duality is a consequence of the exact duality \cite[eq. (5.10)]{Fo09}, proved for $\beta$ even. 
The asymptotic form (\ref{22a}) can readily be checked to be consistent with (\ref{23x}).

\subsection{Relationship to Dyson's log-gas calculation of $E_{\beta}^{\bulk} (n;(0,t))$}

We have seen in the above calculation of the hard edge gap probability $E_{\beta}^{\hard} (n;(0,t))$ 
that expressions obtained in Dyson's log-gas calculation of the bulk gap probability $E_{\beta}^{\bulk} (n;(0,t))$ appear. 
This can be anticipated by noting that under the change of variables $\lambda_l \mapsto \lambda_l^2$, 
the Laguerre $\beta$-ensemble $\mathrm{ME}_{\beta, N} (\lambda^{\beta a/2} e^{-\beta \lambda/2})$ 
transforms to the chiral $\beta$-ensemble specified by a probability density function proportional to
\begin{align}
\nonumber \prod_{l=1}^{N} e^{-\beta \lambda_l^2/2} \lambda_l^{\beta (a'-1/\beta)} \prod_{1\leq j< k \leq N} |\lambda_k^2 - \lambda_j^2|^{\beta}.
\end{align}
The latter has the interpretation \cite[Prop 3.1.4]{Fo10} of a log-gas confined to the half-line $\lambda_l >0$, 
but with image charges of the same sign for $\lambda_l< 0$, and a fixed charge of strength $(a-1)/2$ at the origin.

In the scaled limit of this model the density at the edge $x=0$, as for the bulk, is a constant to leading order, which here takes on the value $1/\pi$ 
(see \cite[eq.~(7.76)]{Fo10}). Consequently, after conditioning so that there are $n$ eigenvalues in $(0,b)$, the task is to compute both $\phi(x)$ and $\rho_{(1)} (x)$ in the equation
\begin{align}
\nonumber \phi(x) = -\int_{0}^{\infty} \left( \rho_{(1)}(y) - 1/\pi \right) \log |x^2-y^2| \; dy.
\end{align}
But defining $\rho_{(1)} (-y) = \rho_{(1)}(y)$, this can be rewritten
\begin{align}
\label{24.1} \phi(x) = -\int_{-\infty}^{\infty} \left( \rho_{(1)}(y) - 1/\pi \right) \log |x^2-y| \; dy.
\end{align}
We observe that with $x^2 \mapsto x$, (\ref{24.1}) is precisely the equation for the potential in the bulk log-gas with background charge density $-1/ \pi$, 
after condition so that there are $2n$ eigenvalues in $(-\sqrt{b}, \sqrt{b})$. This then explains, upon the change of variables $x\mapsto x^2$, 
the appearance of quantities from \cite{Dy95} in our calculation above.

\section{The soft edge}
\setcounter{equation}{0}

\subsection{The case $\beta=2$}

In our log-gas formalism, we are specifying the soft edge by
\begin{align}
\nonumber \rho_b (x) = \frac{\sqrt{x}} {\pi}, \qquad x>0.
\end{align}
In the realization of the soft edge as the scaled neighbourhood of the largest eigenvalue of the Gaussian and Laguerre $\beta$-ensembles
as discussed in the Introduction, this corresponds to shifting the origin to the location of the largest eigenvalue, and then changing the sense of
direction by the mapping $x \mapsto -x$. Thus $x> 0$ now corresponds to the region of the eigenvalue support.
With the system conditioned so that exactly $n$ eigenvalues lie in the interval $(0,t)$, for $1\ll n \ll t$, 
in contrast to the hard wall case we now expect these eigenvalues to have support $(b_1, b_2)$ with $b_1>0$, 
and both endpoints $b_1$ and $b_2$ exhibiting soft edge singularities. We remark that this scenario has recently been
exhibited in Monte Carlo simulations of the eigenvalue profile for the constrained Gaussian $\beta$-ensemble --- the constraint being that a fraction $c > 1/2$ of the
eigenvalues are positive; see \cite[Fig.1 for $x<0$]{MNSV11}. Like the region $x>t$, which we take to have zero potential, 
this region will behave like a perfect conductor and thus we will have $\phi(x) = -v$, $v>0$ for $x\in (b_1, b_2)$.

In this setting (\ref{V1s}) reads
\begin{align}
\label{25} V_1 = -\frac{vn} {2}+ \frac{1} {3\pi} \left( \int_0^{b_1} + \int_{b_2}^t \right) x^{\frac{3}{2}} \frac{d\phi} {dx} dx,
\end{align}
where
\begin{align}
\label{25a} v=\int_{b_2}^t \frac{d\phi} {dx} dx.
\end{align}
To compute $d\phi /dx$ we must generalize the $\alpha =1/2, n=1$ case of (\ref{14.1}). With $b_1$ and $b_2$ now new soft edge singularities, we trial
\begin{align}
\label{25.1} E(z) = i \left( \sqrt{z} - \frac{\sqrt{z-b_1} \sqrt{z-b_2}} {\sqrt{z-t}} \right).
\end{align}
Observe that for this to fall off faster than $z^{-1}$ for $|z| \to \infty$ as required by (\ref{Ezx}), we must have
\begin{align}
\label{25.2} t=b_1+ b_2.
\end{align}
Furthermore, the necessary property (\ref{Ez4}) (with $c=1/\pi, \alpha=1/2$) is satisfied, and $z=t$ is a hard edge singularity as required. 
The real part of (\ref{25.1}) vanishes on the real axis for $x=(b_1, b_2) \cup (t, \infty)$, in keeping with these regions being perfect conductors, 
and use of (\ref{13.2}) shows that $\rho_{(1)} (x)=0$ for $x\in (0, b_1)\cup (b_2, t)$ which again is in keeping with the set up of the problem. 
Finally, we observe that when $n=0$, and thus $b_1 = b_2 = t/2$ (the latter equality follows from (\ref{25.2})), (\ref{25.1}) reduces to (\ref{14.1}) in the case $\alpha=1/2, n=1$. 
Thus (\ref{25.1}) passes all consistency checks, and so is presumed correct in general.

We now make use of (\ref{25.1}) by substitution in (\ref{11.0}) to deduce that
\begin{align}
\label{26.1} \frac{d\phi} {dx}= \begin{cases}
-\sqrt{\displaystyle \frac{(b_1 - x)(b_2 - x)} {t-x}}, &x\in (0, b_1)\\
 \sqrt{\displaystyle\frac{(x- b_1) (x- b_2)} {t-x}}, & x\in (b_2, t)
\end{cases} ,
\end{align}
and by substituting in (\ref{13.2}) to deduce that
\begin{align}
\label{26.2} \rho_{(1)} (x) =\frac{1}{\pi}\sqrt{\frac{(x-b_1) (b_2-x)} {t-x}}, \qquad b_1 < x< b_2.
\end{align}
If we now substitute (\ref{26.2}) in (\ref{d8o}) with $J' = (b_1, b_2)$ and set
\begin{align}
\label{bd} b_1 = t/2 -d, \qquad b_2 = t/2 +d ,
\end{align}
so as to satisfy (\ref{25.2}), we see that $d$ is determined by the requirement that
\begin{align}
\label{27.1} \frac{1}{\pi} \int_{b_1}^{b_2}dx\; \sqrt{\frac{(x-b_1) (b_2- x)} {t-x}} = n.
\end{align}
This integral has the elliptic integral evaluation
\begin{equation}
   n = \frac{2}{3\pi}\sqrt{b_2}\left[ t E'-2b_1K' \right], \quad k^2 = \frac{b_1}{b_2} .
\end{equation} 
For $d/t \ll 1$ this means that $ k \sim 1 $ and the small argument expansions of $ E' $, $ K' $ give that in this limit
\begin{align}
\label{27.1a} d^2 \sim \sqrt{2t} n.
\end{align}

Next, substituting the second case of (\ref{26.1}) in (\ref{25a}) 
\begin{align}
\label{27.2} v := \phi(t)-\phi(b_2) = \int_{b_2}^{t}dx\; \sqrt{\frac{(x-b_1)(x-b_2)} {t-x}} .
\end{align}
This has the evaluation 
\begin{equation}
  v = \frac{2}{3}\sqrt{b_2}\left[ tE-(b_2-b_1)K \right], \quad k^2 = \frac{b_1}{b_2} .
\end{equation} 
We seek the expansion of this integral for $ k^2 \to 1^{-} $ or $ d \to 0 $, and making use of the standard
asymptotic formulae we have
\begin{equation}
  v \sim \frac{1}{3} \sqrt{2} t^{3/2}-\left[1+\log\left(\frac{16 t^2}{d^2}\right)\right]\frac{d^2}{2\sqrt{2t}} .
\end{equation}
Applying this, and making use too of (\ref{27.1a}), we obtain our sought expansion
\begin{align}
\label{29.1} v \sim \frac{4}{3} \left( \frac{t}{2}\right)^{3/2}\left( 1+ \frac{3}{2} \frac{\sqrt{2t} n}{t^2} \log \frac{1}{|t|^{3/4}} \right) .
\end{align}

Presently we have available the leading form of $d$, which according to (\ref{bd}) parameterizes the endpoints of the support of the $n$ eigenvalues, 
in terms of t (the endpoint of the prescribed interval) and $n$, as well as the leading two terms of the potential $-v$ in $(b_1, b_2)$, in terms of $t$ and $n$.
 We next consider the electrostatic potential $V_1$. Substituting (\ref{26.1}) in (\ref{25}) shows
\begin{align}
\nonumber V_1 &= -\frac{vn}{2} +\frac{t}{6\pi} \int_0^{t/2-d}dx\; (t-2x)\sqrt{\frac{(t/2-x)^2 - d^2}{x(t-x)}} ,\\
\label{30} &= -\frac{vn}{2} +\frac{t^3}{6\pi} \int_0^{1-2d/t}dx\; (1-x)\sqrt{\frac{(1-x)^2 - (2d/t)^2}{x(2-x)}}.
\end{align}
As in the case of the hard edge (\ref{19.2}), the integral in (\ref{30}) can be evaluated explicitly.

\begin{lemma}
\label{Lem2}
Let
\begin{align}
\nonumber H(u) := 2\int_0^{1-u}dx\; (1-x)\sqrt{\frac{(1-x)^2 - u^2}{x(2-x)} } .
\end{align}
We have
\begin{align}
\label{30.1} H(u) = \frac{\pi}{2} (1-u^2).
\end{align}
\end{lemma}

\textit{Proof}: Changing variables $x\mapsto 1-x$, then changing variables $x^2=y$ we obtain
\begin{align}
\nonumber H(u) = \int_{u^2}^1 \sqrt{\frac{y-u^2}{1-y}} dy.
\end{align}
Changing variables $y\mapsto y+ u^2$ in this allows the $u$ dependence to be scaled to give
\begin{align}
\nonumber H(u) = (1-u^2) \int_0^1 \sqrt{\frac{y}{1-y}} dy,
\end{align}
and (\ref{30.1}) follows. \hfill $\Box$

Using the result of Lemma \ref{Lem2} in (\ref{30}) gives the evaluation
\begin{align}
\label{V13} V_1= -\frac{vn}{2} +\frac{t^3}{24} \left( 1- \frac{4d^2}{t^2} \right).
\end{align}
Substituting (\ref{27.1a}) and (\ref{29.1}) it follows from this that for $1\ll n\ll t$ and $t \to \infty$
\begin{align}
\label{29a} V_1 \sim \frac{t^3}{24} - \frac{t\sqrt{2t}}{3}n -\frac{1}{2} n^2 \log \frac{1}{t^{3/4}}.
\end{align}
Since the term (\ref{V2}) vanishes for $\beta=2$, for this $\beta$ we can substitute (\ref{29a}) for $\delta F$ in (\ref{D6}) to include
\begin{align}
\label{29b} \log E_2^{\soft} (n;(t,\infty)) \mathop{\mathop{\sim}\limits_{t\to -\infty}}\limits_{1\ll n\ll |t|}
 -\frac{|t|^3}{12} + \frac{|t|^{3/2} 2\sqrt{2} n}{3} +n^2 \log \frac{1}{t^{3/4}}.
\end{align}
Although there is no exact asymptotic results in this regime, as with (\ref{22}) the exact expansion with $\beta=2$ and $n$ fixed is known \cite{TW94a}, 
\cite[eq. (9.89)]{Fo10}, and its large $t$ expansion is in complete agreement with (\ref{29b}).

\subsection{$V_2$ in the case $n=0$ and general $\beta > 0$}

In the case $n=0$, $\rho_{(1)}(x)$ is supported on $(t, \infty)$ and according to (\ref{25.1}) with $b_1=b_2 =t/2$ substituted in (\ref{13.2}) takes on the explicit functional form
\begin{align}
\nonumber \rho_{(1)}(x) = \frac{x-t/2} {\pi\sqrt{x- t}}.
\end{align}
Substituting in (\ref{V2}) and a simple change of variables shows 
\begin{align}
\nonumber V_2 = \left( \frac{1}{\beta} - \frac{1}{2} \right) \frac{t^{3/2}}{\pi} \int_1^{\infty} \frac{x-1/2}{\sqrt{x-1}} \; \log \frac{x-1/2}{\sqrt{x} \sqrt{x-1}} \; dx.
\end{align}
This integral can be evaluated by computer algebra to give
\begin{align}
\label{30t} V_2 = \left( \frac{1}{\beta} - \frac{1}{2}\right) \frac{t^{3/2}}{3} (\sqrt{2} -1/2).
\end{align}
Adding this to (\ref{14.2a}) with $c=1/\pi$, then substituting for $\delta F$ in (\ref{D6}) implies
\begin{align}
\label{30y} \log E_{\beta}^{\soft} (0;(t,\infty)) \mathop{\sim}\limits_{t\to -\infty} -\frac{\beta |t|^3}{24} + \left( \frac{\beta} {2} -1\right) \tilde{k} |t|^{3/2}
\end{align}
with $\tilde{k}= (\sqrt{2} -1/2)/3$. However, this contradicts the known \cite{BBD07, BEMN10} $t\to -\infty$ asymptotic form of $E_{\beta}^{\soft}(0;(t,\infty))$, 
which tells us that (\ref{30y}) holds with $\tilde{k}=\sqrt{2}/3$. Thus, as for the hard edge, 
 this infinite system implementation of the log-gas formalism breaks down in its form of $V_2$. But we observe that the correct value of $\tilde{k}$ follows
 if we replace $V_2$ by  (\ref{V2v1}).

\subsection{$V_2$ for general $1\ll n\ll t$ and general $\beta > 0$}

The integration method used by Dyson in \cite{Dy95} to deduce (\ref{V2v}) allows us to deduce that 
\begin{align}
\label{31x} V_2 = \left( \frac{1} {\beta}- \frac{1}{2} \right) \left[ \frac{1}{2}v + \frac{1}{2}(\phi(b_1)-\phi(0)) \right] .
\end{align}
We don't give the details since we have already established that this term does not correctly reproduce known exact results.

Instead, we proceed by analogy with the bulk, the corrected hard edge result  and the corrected $n=0$ soft edge result, and thus hypothesize that (\ref{31x}) should be replaced by (\ref{V2v1}).
Doing this, deducing the asymptotic form of $V_2$ using (\ref{29.1}) and adding it to (\ref{29a}), then gives 
\begin{multline}
\log E_{\beta}^{\soft} (n; (t,\infty)) \mathop{\mathop{\sim}\limits_{t\to -\infty}}\limits_{1\ll n\ll |t|} 
  -\frac{\beta |t^3|} {24} + \frac{\sqrt{2}}{3} |t|^{3/2} \left( \beta n + \frac{\beta}{2} -1 \right)\\
\label{31g} 
  +\left[ \frac{\beta}{2}n^2 + \left( \frac{\beta}{2} - 1\right)n \right] \log |t|^{-3/4}.
\end{multline}
As for $E_{\beta}^{\hard} (n;(0,t))$, in the case $n=0$ the form of the term $O(\log |t|)$ in the $t\to -\infty$ asymptotic expansion of $\log E_{\beta}^{\hard} (n;(0,t))$ is known, 
and this added to (\ref{31g}) suggests that uniformly in $n$, for $n\ll t$, the expansion (\ref{31h}) as recorded in the Introduction holds true.

Again as for the hard edge case, there is an asymptotic functional relation which (\ref{31h}) must satisfy \cite{Fo09}. This states
\begin{align}
\label{32z} \log E_{\beta}^{\soft} (n; (s_{\beta} t, \infty)) \mathop{\mathop{\sim}\limits_{t\to -\infty}}\limits_{n\ll t} 
\log E_{4/ \beta}^{\soft} \left( \beta n/2+\beta/2 -1; (s_{4/ \beta}t, \infty) \right),
\end{align}
where $s_{\beta}, s_{4/\beta}$ are length scales related by $(\beta/2)^{2/3} s_{\beta} = s_{4/\beta}$, and this is indeed exhibited by (\ref{31h}). 
It is similarly true that (\ref{31h}) is consistent with the soft edge analogue of (\ref{22.1}) \cite{Fo06c} 
\begin{align}
\nonumber E_2^{\soft} (n; (t,\infty)) \mathop{\mathop{\sim}\limits_{t\to -\infty}}\limits_{n\ll |t|} E_1^{\soft} (n; (t,\infty)) E_1^{\soft} (n+1; (t,\infty)).
\end{align}

\section{Concluding remarks}\label{sVs}
An infinite log-gas formalism, due in the bulk to Dyson \cite{Dy95} and independently Fogler and Shklovskii \cite{FS95}, has been applied to the
computation of the conditioned soft and hard edge gap probabilities $E^{\soft}_{\beta} (n; (t,\infty))$ and $E^{\hard}_{\beta} (n; (0,t))$. For this purpose
the hard and soft edges are characterized by their asymptotic densities (\ref{hardd}) and (\ref{softd}), which are taken to be the exact profiles of
the background densities in the log-gas. The hypothesis (\ref{D6}) asserts that the conditioned gap probability is given in terms of  the change in free energy
resulting from the conditioning. In the bulk, it was found in  \cite{Dy95}  and \cite{FS95} that the first three orders in the large $t$ expansion of
$E^{\rm bulk}_\beta(n;(0,t))$, with $0\ll n \ll t$ could be obtained by asserting that the change in free energy is due to an electrostatic and entropy
term. The latter happens to be simply related to the potential drop in going from the bulk of the Coulomb fluid, to the region containing the conditioned
$n$ eigenvalues (recall (\ref{V2v1}) above).

At the hard and soft edges, it is found that following the same ansatz as used in  \cite{Dy95}  and \cite{FS95} leads to an inconsistency with known
asymptotic results for the $n=0$ case of $E^{\soft}_{\beta} (n; (t,\infty))$ and $E^{\hard}_{\beta} (n; (0,t))$. This inconsistency is due to the entropy term.
On the other hand, if we use instead  (\ref{V2v1}), which in the bulk is identical with the entropy term, consistency with the known $n=0$ results is obtained.
Making this replacement for general $0 \ll n \ll t$ allows us to make the asymptotic predictions (\ref{22a}) and (\ref{31h}). 

The infinite log-gas approach should be contrasted to Dyson's pioneering strategy  \cite{Dy62b}, applied to the computation of the asymptotic formula
 (\ref{Eb})
for $E^{\rm bulk}_\beta(n;(0,t))$ with $n=0$. The calculation consists of two steps: first use the log-gas hypothesis (\ref{D6}) to compute the large $N$ form of the  probability that there are no eigenvalues in the angular interval $[0, \alpha]$ in the circular $\beta$-ensemble.
This is a large deviation quantity as there would ordinarily be on average $\alpha N/2\pi$ eigenvalues in this interval. Second scale the variable $\alpha$
in this large deviation formula by writing $\alpha = 2 \pi t/N$ to deduce (\ref{Eb}).
It was over 30 years later that the log-gas argument was extended to the infinite system \cite{Dy95, FS95}, 
and applied to the study of $E_{\beta}^{\bulk} (n; (0,t))$.

Large deviation log-gas studies of the soft edge of the Gaussian $\beta$-ensemble were initiated by Dean and Majumdar \cite{DM06, DM08},
and by Vivo, Majumdar and Bohigas \cite{VMB07} at the soft edge of the Laguerre ensemble.
By an appropriate scaling the first term in (\ref{30y}) was deduced. In these works, only the electrostatic contribution $V_1$ was considered. 
More recently \cite{BEMN10}, using a formalism based on topological recursions pioneered by Eynard and collaborators, 
higher order correction terms to the soft edge gap probability of the Gaussian $\beta$-ensemble were obtained 
(see \cite[eq.~(20)]{NM11} for its explicit functional form). Appropriate scaling then gave the expansion (\ref{31h}) in the case $n=0$, 
together with the explicit form of the next term in the asymptotic expansion which is a $\beta$-dependent constant (in relation to the latter, see also \cite{BBD07,DIK08}).
 The significance of this from the log-gas perspective is that the first two terms of the topological recursion are precisely $V_1$ and $V_2$ from the log-gas formalism. 
In particular, unlike our finding for the infinite log-gas approach at the soft edge, 
the large deviation log-gas computation does correctly give the first two terms in the asymptotic expansion. 

This same conclusion holds true at the hard edge. 
The work of Chen and Manning \cite{CM94}, applying what can be viewed as a large deviations  log-gas argument  at the hard edge of the Laguerre ensemble, 
then performing a double scaling limit to get the asymptotics of the hard edge gap probability correctly gives terms the first two (and the third also) terms.

The question then remains as to why the entropy term $V_2$ in the double scaling limit of the large deviation approach leads to correct results for the
asymptotic forms of $E^{\soft}_{\beta} (n; (t,\infty))$ and $E^{\hard}_{\beta} (n; (0,t))$ with $n=0$, whereas in the infinite log-gas approach this term must be
replaced by  (\ref{V2v1}).  And why in the bulk  (\ref{V2v1}) is an identity for $V_2$ so no modification is required. One possible explanation for the
breakdown of the form  (\ref{V2})  for $V_2$ at the edge but not in the bulk relates
to the approach to the edge and bulk states from a finite system. The log-gas on a circle has uniform density for all $N$ and so approximates the bulk without
any boundary term corrections. In contrast, the hard and soft edge states for finite $N$, as distinguished by the density profile, are only local regions of the
spectrum. But the entropy term $V_2$ (\ref{V2}) is a global quantity, and carries information about the global spectrum which is lost if starting from
a forever extending edge state as in the infinite log-gas formalism. If we accept this, the remaining question, on which we have no immediate answer,
is to explain the explicit form of the corrected formula for $V_2$, (\ref{V2v1}).

\subsection*{Acknowledgements}
This work was supported by the Australian Research Council. The assistance in the preparation of this manuscript by Anthony Mays is acknowledged, as
are the considered remarks of the referee.


\providecommand{\bysame}{\leavevmode\hbox to3em{\hrulefill}\thinspace}
\providecommand{\MR}{\relax\ifhmode\unskip\space\fi MR }
\providecommand{\MRhref}[2]{%
  \href{http://www.ams.org/mathscinet-getitem?mr=#1}{#2}
}
\providecommand{\href}[2]{#2}

\end{document}